\documentclass[conference]{IEEEtran}
\IEEEoverridecommandlockouts
\usepackage{cite}
\usepackage{amsmath,amssymb,amsfonts}
\DeclareMathOperator*{\argmax}{argmax} % thin space, limits underneath in displays
\usepackage{algorithmic}
\usepackage{graphicx}
\usepackage{textcomp}
\usepackage{mathtools}
\usepackage{caption}
\usepackage{float}
\usepackage{xcolor}
\usepackage{lineno}
\usepackage{booktabs}
\def\BibTeX{{\rm B\kern-.05em{\sc i\kern-.025em b}\kern-.08em
    T\kern-.1667em\lower.7ex\hbox{E}\kern-.125emX}}

\begin{document}

\title{A Novel ECG Signal Denoising Filter Selection Algorithm Based on Conventional Neural Networks\\
}

\author{\IEEEauthorblockN{Chandresh Pravin}
\IEEEauthorblockA{{Department of Computer Science,} \\
{University of Reading,}\\
{Reading, United Kingdom} \\
0000-0003-1530-0121}
\and 
\IEEEauthorblockN{Varun Ojha}
\IEEEauthorblockA{{Department of Computer Science,}\\
{University of Reading,} \\
{Reading, United Kingdom} \\
0000-0002-9256-1192}
}

\maketitle

\begin{abstract}

%We propose a deep learning based denoising filter selection method for Electrocardiograph (ECG) signal preprocessing. ECG signals obtained under clinical conditions, such as those measured using skin contact devices in hospitals, often contain noise and baseline signal disturbances, for signals acquired under non-clinical conditions in fact, such as heart rate signatures acquired using non-contact ultra-wideband radar systems, they consist of relatively greater levels of noise than clinical signals. This is mainly attributed to the measuring equipment used and the signal processing techniques applied in both signal acquisition techniques. Often, the noise present in different signal waveforms exhibit minute variances that require different processing techniques for effectively denoising such signals. The overall task of reducing noise from an ECG signal is to make the analysis of its waveform characteristics more accurate, be it through manual or computational examination. Currently, the most common method of filtering noise from ECG signals is through the use of wavelet filters. There are, however, circumstances in which using a different filtering method may result in higher signal-to-noise-ratios (SNR) for a waveform; in this paper, we investigate wavelet filtering and elliptical filtering methods for the task of reducing noise in signals. Our proposed convolutional neural network architecture classifies (with 92.8\% accuracy) the optimum filtering method for noisy signal based on its expected SNR value.

We propose a novel deep learning based denoising filter selection algorithm for noisy Electrocardiograph (ECG) signal preprocessing. ECG signals measured under clinical conditions, such as those acquired using skin contact devices in hospitals, often contain baseline signal disturbances and unwanted artefacts; indeed for signals obtained outside of a clinical environment, such as heart rate signatures recorded using non-contact radar systems, the measurements contain greater levels of noise than those acquired under clinical conditions. In this paper we focus on heart rate signals acquired using non-contact radar systems for use in assisted living environments. Such signals contain more noise than those measured under clinical conditions, and thus require a novel signal noise removal method capable of adaptive determining filters. Currently the most common method of removing noise from such a waveform is through the use of filters; the most popular filtering method amongst which is the wavelet filter. There are, however, circumstances in which using a different filtering method may result in higher signal-to-noise-ratios (SNR) for a waveform; in this paper, we investigate the wavelet and elliptical filtering methods for the task of reducing noise in ECG signals acquired using assistive technologies. Our proposed convolutional neural network architecture classifies (with 92.8\% accuracy) the optimum filtering method for noisy signal based on its expected SNR value.

\end{abstract}


\section{Introduction}
With the emergence of embedded devices capable of artificial intelligence and learning abilities, there remains the problem of effective data processing, which has been a task for engineers and data analysts alike since the development of embedded systems entirely~\cite{KHOKHAR20151650}. Acquiring (in the form of signals) and analysing (signal processing) information from the natural phenomena that occurs in the real-world comes with various difficulties; none more so prevalent than the task of differentiating between information that is important for analysis, from that which is not. Signals acquired using assistive technologies from applications outside of a laboratory setting are inherently subjected to greater levels of signal fluctuations from the recording environment. These fluctuations, also considered as noise, are often detrimental to the analysis of the objective event. The role of a digital signal processor, for the majority of the data used in modern applications is considered electronically, is to adequately remove unwanted artefacts or disturbances from the recorded signals that represent some target real-world event. We propose a Deep Neural Network (DNN) architecture to recognise minute variations of an objective signal and classify the optimum method for removing noise from the waveform. The objective application of this study is to outline an effective and automatic denoising method for Electrocardiograph (ECG) signals measured using ultra-wideband (UWB) radar systems to be used in assisted living environments. Measurements made using UWB radar systems use micro-Doppler signatures to identify subtle movements from an individual's body, such as heart rate and respiration rate. Due to the unobtrusive method of data acquisition, recorded signals are often contaminated with varying levels of noise from different sources that need to be removed in order to analyse the vital signs effectively~\cite{paulson2005ultra}, we propose an automated method of removing noise from inherently noisy ECG signal recordings.

Oftentimes the disparity between studies carried out under clinical environments and the technology being used in real-world applications stem from the attention paid to the data being considered, or there is a lack of for the later condition. Classical machine learning models and more recently, deep learning models, have already made significant strides in recognising important information from noisy and often weak signals~\cite{Nassif2019Speech, Foster2014MachineLM}. The need for an automated model to recognise random, seemingly unpredictable variances in the environment is therefore needed to understand how best to process raw signals. For this study, the signals considered as the base ECG waveforms are from the MIT-BIH normal sinus rhythm database~\cite{moody2001impact}. In this paper, we propose a model to predict the best method for removing unwanted artefacts from a noisy ECG waveform. In regular applications, disturbances in a recorded waveform are removed using various signal processing techniques; methods primarily based around filtering-out unwanted noise and retaining the important features from the signal. The task of filtering a raw signal involves determining the optimum filter, along with other hyperparameters, such that the filter effectively removes the noise present in the signal. This requires careful consideration and analysis of the data at hand, often in the frequency domain. Furthermore, static methods of filtering, where the system function does not adapt in behaviour to changes in the input, are at risk of attenuating important features from the waveform, or equally, being unable to sufficiently remove certain aspects of noise, both of which may affect further analysis of the evaluated signal~\cite{proakis2007digital}. 

Signals measured outside of a laboratory, such as assisted living spaces where environmental factors are seldom controlled, the acquired waveforms are intrinsically noisy and irregular, thus such signals require a filtering method that is able to adapt to these variances. Haykin details the process of adaptive filtering~\cite{Haykin1996Adaptive} where an optimisation algorithm is used to determine the adjustable filter parameters, however, such a method is still frequently inadequate in dealing with subtle deviations in noise signal, particularly where the noise in a signal varies regularly, requiring a more stochastic approach to filtering noise~\cite{kovac2013adaptive}. Deep learning models in contrast, have the innate capacity for recognising detailed patterns, and such, make them ideal for differentiating signals by recognising minute differences between the waveforms. The intended real-world application for such a model; being used in assistive technologies used to remotely monitor the cardiac health of individuals in an environment, introduces a constraint of signal windowing in order to reduce the number of elements to be considered at a given time, such that the signals may be analysed in real-time. This constraint consequently also reduces computational complexity of the overall process~\cite{Joshi2013ASurvey}. 

In section~\ref{sssec:results}, we detail how commonalities can be found between different signals that result in higher SNR values for a given filter, thus also outlining the training process of the proposed model. The architecture outlined in this study is that of a binary classifier intended to predict an optimum filtering method, between wavelet filtering and elliptical filtering, for a noisy input signal. The Elliptical filter was chosen as an alternative to the wavelet filter due to its narrow transition response at the cut-off frequency, respective to other similar finite impulse response (FIR) filter functions, such as the Chebyshev and Butterworth filters~\cite{proakis2007digital}. After classifying the optimum filter, a newly presented waveform is filtered using the filter label and predefined filter coefficients, which are determined during the model training process.

The paper is organised as follows: Section~\ref{sssec:related_word} contains a review of relevant techniques and methods explored in this paper, section~\ref{ssssec:methods} outlines a definition of the signals used~\ref{subsec:sig_def}, model training algorithm ~\ref{subsec:model_training}, classification model parameters~\ref{subsec:model_training}, feature reduction techniques used~\ref{subsec:feature_reduction} and finally the experimental set-up~\ref{subsec:experimenta_setup}. This is followed by the results and discussions section~\ref{sssec:results}, which details the experimental results~\ref{subsec:experimental_res} and a discussion of the results~\ref{subsec:discussion}. Finally, concluding remarks on the study carried out and possible future avenues of research are presented in~\ref{ssssec:conclusion}.

\section{Related Work}\label{sssec:related_word}

Physiological signals have been recorded and studied extensively for many years and none more so than ECG signals. The analysis of ECG signals is vital in diagnosing, and often treating, the cardiac health of a person; this has until recently, been a task carried out by medical professionals exclusively~\cite{huikuri2001sudden}. With the power of modern machine learning tools and the availability of assistive technologies to monitor, process and analyse ECG signals remotely and in real-time, the task of detecting and predicting cardiac abnormalities have become a task for medical professionals and machine learning engineers alike~\cite{MARTIS201211792}. We focus on the noise present within ECG signals, which varies from one segment of the signal to another and is often unable to be removed effectively with standard filtering methods~\cite{Joshi2013ASurvey}. For such instances, machine learning and deep learning networks have been applied to remove noise in ECG signals~\cite{antczak2018deep, acharya2017application}. %The effectiveness of these models, however, decreases as the noise power within the signal increases, as is the nature signals acquired using UWB radar systems.

With instances of vital signs monitoring using UWB radar system-based assisitive technologies, various investigations have been carried out into ECG signal denoising and classification tasks~\cite{li2013advanced, lazaro2010analysis}. The studies described go as far as providing a holistic overview of the data acquisition process and demonstrating various signal processing techniques to extrapolate vital signs form radar signatures. Similar to ECG detection, Liang et al.~\cite{liang2018ultra} proposed a method of detecting respiration signs using a frequency accumulation algorithm followed discrete short-time-Fourier transform to suppress random signal harmonics and products of heartbeat and respiration signals combined. A model proposed by Shikhsarmasr et al.~\cite{Shikhsarmast_2018} makes use of the wavelet packet decomposition method to suppress random noise in the signal, subsequently makes use of a vital sign estimation model on a defined region of interest, to improve the overall system efficiency. The wavelet transform has shown to be used in many similar studies~\cite{kabir2012denoising,POUNGPONSRI2013206}, whereby a thresholding method is used to attenuate certain frequency amplitudes of the signal noise.

There exist studies using deep learning and machine learning algorithms in various forms to carry out the task of ECG signal denoising. Antczak~\cite{antczak2018deep}, proposed using synthetic data to train a deep algorithm for signal denoising and fine-tuning the network parameters to learn higher-level features using real data. More recently, a layer-by-layer denoising neural network has been developed based on factor analysis~\cite{wang2020ecg}, where the model attempts to learn Gaussian noise present in the signal, thus being able to remove it. A similar study has been carried out for an audio equalisation task by Pepe et al.~\cite{Pepe_2020}, in which, given a noisy signal, the FIR filter coefficients are predicted using a DNN architecture. Indeed, various attempts have been made into developing learning models capable of determining optimum filtering coefficients for reducing noise in ECG signals~\cite{POUNGPONSRI2013206, Jenkal2016499}. To the best of our knowledge through a review of relevant literature we for the first time propose a novel study detailing the use of machine learning techniques for the task filter selection in ECG signal denoising. 

%The novelty of the model proposed in this research stems from its application; developing a model using a DNN to classify the optimum denoising filtering method, based on prior training.

\section{Denoising Filter Classification Modelling}\label{ssssec:methods}

The primary application for the model proposed in this study is to find the best method of attenuating unwanted artefacts from a noisy ECG signal acquired using an UWB radar system. To simulate additional noise to the signals, as would be expected from radar readings of this nature~\cite{ramirez2001performance}, Gaussian noise was added to the original dataset. The new signals were then processed using a wavelet filter and a low-pass elliptical filter to determine the optimum method for reducing the noise in a signal. Finally, a classification model was trained and tested using the labelled dataset for the purpose of predicting the filtering method which results in the highest SNR value.

\subsection{ECG Signal Definition}\label{subsec:sig_def}

A raw ECG signal, being a continuous-time signal, can be viewed, through sampling as a discrete time signal \(x(k)\) as per the following definition:
\begin{equation}
\label{eq:signal_def}
    x(k) \triangleq x(t)\mid {t=kT}, 
\end{equation}
where \(k = \{1,\ldots,K\}\) and represents the number of discrete data points in the waveform. The parameter \(T\) being the sampling period of the discrete-time signal and thus the sampling frequency is given as $\nu_s = 1 / T$. The raw is represented by $x(k)$, however, having been originally recorded under clinical conditions, can be regarded as the optimum ECG waveform for this investigation. This primary aim of this study is to improve the signal quality of ECG signals recorded using UWB radar systems, which exhibit greater levels of noise than the raw signal used in this study.

To modify the raw signal in order to make it more noisy, artificial noise $g(k)$ is generated using the Gaussian distribution function as per: 
\begin{equation}
\label{eq:gauss_func}
    g(k) = \frac{1}{\ \sigma\sqrt{2 \pi}} e^{ - \frac{ (r(k) - \mu)^2 } {2 \sigma^2}},
\end{equation}
% https://www.itl.nist.gov/div898/handbook/eda/section3/eda3661.htm
%
where $r(k)$ is a random value between 0 and 1 and  \(\mu\) and \(\sigma\), are the mean and standard deviation of the signal $x(k)$. The noise being modelled on the characteristics of the original signal itself. The waveform to be considered by the classifier is $z(k)$, being a noisy ECG signal which is generated by applying the original signal $x(k)$ with the generated Gaussian distributed noise $g(k)$:

\begin{equation}
\label{eq:signal}
    z(k) = x(k) + g(k).
\end{equation}
The objective of the classification model is to determine the optimum method to reduce the noise $g(k)$, from the noisy signal $z(k)$.

\subsection{Filter Label Identification Algorithm}\label{subsec:model_training}

The resultant signal $z(k)$ is firstly normalised using min-max normalisation and subsequently processed through an SNR optimisation function  $\Omega(\cdot)$ that returns signal label $y$ defined as a composition of the filter label, $\alpha$; maximum SNR value, $\beta$; and the chosen filter variable value $\delta$, as shown in~\eqref{eq:snr_optl}. We reshape the clean signal $x(k)$ and noisy signal $z(k)$ of lengths $K$ into a $M \times N$ shaped matrices as follows:

\begin{equation}
\label{eq:signal_transform}
   \text{X} \in \mathbb{Q}^{M \times N}, \text{Z} \in \mathbb{Q}^{M \times N} ~|~ \mathbb{Q}^{M \times N} \leftarrow  \mathbb{Q}^{1 \times K},  \\
\end{equation}

\noindent where $\text{X} = [\textbf{x}^1, \textbf{x}^2, \ldots, \textbf{x}^{M} ]^T$ and $\text{Z} = [\textbf{z}^1, \textbf{z}^2, \ldots, \textbf{z}^{M} ]^T$ are the clean and noisy signals reshaped into a matrix of windowed signals. The reshaping parameters $M$ and $N$ are non-zero natural numbers defined as \(M=K/\lambda\) and \(N = \lambda\), where  $ \lambda $ is the chosen window period. The variables \(m\) and \(n\) are the matrix indices and are natural numbers, such that \(m = \{1,2,\ldots,M\}\) and \(n = \{1,2,\ldots,N\}\). This process is carried out to reshape the original signal of length $K$ into $M$ windowed signals of length $N$, an example for $\text{Z}$ is shown in the following form:

\begin{equation}
\label{eq:sig_matrix}
    \text{Z} = \left[
    \begin{array}{c}
    \textbf{z}^1 \\
    \textbf{z}^2 \\
    \vdots\\
    \textbf{z}^M\\
    \end{array}
    \right]
    \equiv
    \left[
    \begin{array}{ccccc}
       z^1_1  & z^1_2 & \ldots & z^1_{N} \\
       z^2_1  & z^2_2 & \ldots & z^2_{N} \\
       \vdots &\vdots & \ddots & \vdots \\
       z^{M}_1 & z^{M}_2 & \ldots & z^{M}_{N} \\
    \end{array}
    \right]~,
\end{equation}

\noindent where the windowed noisy signal is written as $\textbf{z}^m = (z^m_1,z^m_2,\ldots,z^m_{N})$ and equally the windowed clean signal can be given as $\textbf{x}^m=(x^m_1,x^m_2,\ldots,x^m_{N})$. The function $\Omega(\cdot)$ being applied to the resultant windowed signals returns the corresponding filter labels $y^m$ given as:
\begin{equation}
\label{eq:snr_optl}
    y^m = (\alpha^m, \beta^m,\delta^m) = \Omega(\textbf{x}^m, \textbf{z}^m, \theta_f),
\end{equation}
where $\theta_f$ is the maximum value of the filter boundary parameter and $f$ is the filter label. We assign $f=0$ for elliptical filter and $f=1$ for wavelet filter. The variable $\theta_f$ depends on the filter being used, since the elliptical and wavelet filters have different parameters types. When considering the elliptical filter, $\theta_0$ represents the maximum cut off frequency for a low pass filter. As for the wavelet filter, $\theta_1$ represents the maximum number of wavelets to be investigated and must satisfy \(\theta \leq \Theta\), where $\Theta$ is the maximum number of wavelets available. The clean signal $\textbf{x}^m$, is used within the SNR optimisation function for calculating the individual SNR values of the filtered signals.

The maximum SNR value $\beta_f^m$, as shown in~\eqref{eq:snr_val}, and optimum filter variable $\delta_f^m$ that returns $\beta_f^m$, given by~\eqref{eq:delta_init}, for a given filter $f$ and windowed signal index $m$ is obtained by:

\begin{equation}
\label{eq:snr_val}
    \beta_f^m = \max[\mbox{SNR}(\textbf{x}^m, \textbf{r}^m_f)]
\end{equation}
\begin{equation}
\label{eq:delta_init}
    \delta^m_f = \argmax_{\omega_c}[\mbox{SNR}(\textbf{x}^m, \textbf{r}^m_f)],
\end{equation}
where $\textbf{r}^m_f$ is a signal obtained though a function $R_f(\textbf{z}^m, \omega_c)$ for a filter $f$ and windowed signal with index $m$. The filtered signal $\textbf{r}^m_f$ carries the same length as $\textbf{z}^m$, such that $\textbf{r}^m_f = (r^m_{f,1}, r^m_{f,2},\ldots,r^m_{f,N})$. The critical filter variable is $\omega_c \in \mathbb{N} \text{ and } \theta_{f,0} \ge \omega_c > \theta_f$, where the term $\theta_{f,0}$ holds the initialising value of $\omega_c$. As the elliptical and wavelet filters accept different function parameters, the values of $\omega_c$ and $\theta_{f,0}$ carry different representations for each filter. For the elliptical filter, where $f=0$, the value $\theta_{0,0}$ is the lowest frequency to be tested, which is chosen to be 1Hz. Given the condition where $f=1$, the value of $\theta_{1,0}$ is the first element from an ordered set containing numerically encoded wavelets, such that each element from the set can be decoded to retrieve its corresponding wavelet type. The function $ R_f(\textbf{z}^m, \omega_c)$ is defined as:
\begin{equation}
\label{eq:filter_eq}
    \textbf{r}^m_f = R_f(\textbf{z}^m, \omega_c) = \textbf{T}_f^{-1}[H_f(\omega_c)\cdot \hat{\textbf{z}}^m],
\end{equation}
where $\textbf{T}_f^{-1}$ is the general inverse transform operator, given as the inverse fast Fourier transform (FFT) when $f=0$, and is the inverse discrete wavelet transform (DWT) when $f=1$. This transformation is applied to a convolution of the filter function $H_f(\omega_c)$ and $\hat{\textbf{z}}^m$, being the windowed signal $\textbf{z}^m$ transformed in the Fourier domain given $f=0$ and in the wavelet domain given $f=1$ and $\omega_c$. 

When using the elliptical filter function $H_0(\omega_c)$, the Nyquist theorem must be satisfied before applying the filter, as shown:
\begin{equation}
\label{eq:ellip_eq}
    H_0(\omega_c) = \{\Psi_p(\omega, \omega_0)~|~\omega_0 = \frac{\omega_c}{0.5~\nu_s})\}~,
\end{equation}
\begin{equation}
\label{eq:ellip_psi}
    \Psi_p(\omega, \omega_0) = \frac{1}{\sqrt{1+\epsilon ^2 R_p^2(\xi,\frac{\omega}{\omega_0})}}~
\end{equation}
where $\omega_0$ is the cut-off frequency, $\omega$ is the angular frequency given as $2\pi \nu$ (where $\nu$ is the ordinary frequency in Hz), $\epsilon$ is the ripple factor and $\xi$ is the selectivity factor. The $p$-th-order elliptical filter is indicated by $\Psi_p(\omega,\omega_0)$, and the function $R_p$ referred to as a Chebyshev rational function that controls the stopband ripple response. The ripple factor specifies the passband ripple, whereas the stopband ripple is given by the combination of the ripple factor and selectivity factor.
% https://en.wikipedia.org/wiki/Elliptic_filter

Similarly, an example function $H_1(\omega_c)$ representing a wavelet filer is used (as mentioned in \textit{pywt}\footnote{https://pywavelets.readthedocs.io/en/latest/} library). When we use the wavelet filtering method, we apply DWT using a wavelet, denoted by $\omega_c$, that results in a transformed signal $\hat{\textbf{z}}^m$. Following this, we apply thresholding to the wavelet detail coefficients $u$ using a soft threshold $\hat{u}$, defined in~\cite{zhang2019ecg}, as per:
\begin{equation}
 \label{eq:soft_thres}
    u = 
    \begin{cases}
    [\textit{sgn}(u)](|u| - \hat{u}) & \; |u| \ge \hat{u}\\
    0 & \; |u| < \hat{u}
    \end{cases}
\end{equation}

\begin{equation}
\label{eq:thresholding}
    \hat{u} = \sigma \sqrt{2 \log{N}}~,
\end{equation}
where $N$ and $\sigma$ are the length and standard deviation of the input signal $\textbf{z}^m$ respectively. The noise present in the signal is then attenuated using a thresholding function, given by~\eqref{eq:thresholding}, before reconstructing the original signal using the inverse of DWT given by~\eqref{eq:filter_eq}.

From~\eqref{eq:snr_val} the value of $\beta_f^m$ is given as the maximum SNR value achieved for a windowed signal $m$ being processed through filter $f$ and it can be seen from~\eqref{eq:delta_init} that $\delta_f^m$ takes the value of $\omega_c$ for which $\beta_f^m$ is achieved. The standard SNR function used in both~\eqref{eq:snr_val} and \eqref{eq:delta_init} is given by: 
\begin{equation}
\label{eq:SNR_eq}
    \mbox{SNR}(\textbf{x}^m, \textbf{z}^m) = 10 \log_{10} {\left(\frac{\sum\limits_{n=1}^{N}[{x}^m_n]^2}{\sum\limits_{n=1}^{N}[{x}^m_n-{z}^m_n]^2}\right)}~,
\end{equation}
where $\textbf{x}^m$ is the clean signal and $\textbf{z}^m$ is the signal to be compared. 
The root-mean-square error (RMSE) calculation is used in Section~\ref{subsec:discussion} to compare the average power difference between noisy signals and their corresponding clean signals, as given by~\eqref{eq:rmse_eq}. The definition of RMSE being the standard deviation of the residuals (predicted errors) between two given signals $\textbf{x}^m$ and $\textbf{z}^m$, given by:
\begin{equation}
\label{eq:rmse_eq}
    \mbox{RMSE}(\textbf{x}^m, \textbf{z}^m) = \sqrt{\frac{1}{N}\sum\limits_{n=1}^{N}(x^m_n-z^m_n)^2}~.
\end{equation}

The equivalent values of the window label $y^m$, being the filter label $\alpha^m$, the maximum SNR value $\beta^m$, and the optimum filter variable $\delta^m$, are given as per:
\begin{equation}
\label{eq:alpha_val}
    \alpha^m = \left\{
    \begin{array}{rl}
         0 & \text{if} \; \beta_0^m \ge \beta_1^m\\
         1 & \text{if} \; \beta_0^m < \beta_1^m\\
    \end{array}
    \right.
\end{equation}
\begin{equation}
\label{eq:snr_compare}
    \beta^m =\max\{\beta_0^m, \beta_1^m\}.
\end{equation}
%
% \begin{equation}
% \label{eq:delta_eq}
%     \delta^m = 
%     \begin{cases}
%         \frac{1}{count(\delta_f^m)}\sum\limits_{m=0}^{M}(\delta_f^m)     & \text{if} \; f = 0\\
%          mode(\delta_f^m)   & \text{if} \; f = 1\\
%     \end{cases}
% \end{equation}
%
%%%%%%%%%%%%%%%%%%%%%%%%%%%%%%%%%%%%%%%%%%%%%%%%%%%%%%%%%%%%%%
% \begin{equation}
% \label{eq:delta_eq}
%     \delta^m = 
%     \begin{cases}
%          \delta_0   & \text{if} \; \alpha^m = 0\\
%          \delta_1   & \text{if} \; \alpha^m = 1\\
%     \end{cases}
% \end{equation}
%%%%%%%%%%%%%%%%%%%%%%%%%%%%%%%%%%%%%%%%%%%%%%%%%%%%%%%%%%%%%%
Here the filter label $\alpha^m$ and the maximum SNR value $\beta^m$ are shown by~\eqref{eq:alpha_val} and~\eqref{eq:snr_compare} respectively. Note that the individual signal filter labels are assigned using~\eqref{eq:alpha_val}, depending on which maximum SNR value for a given filter $\beta^m_f$ resulted in the highest overall SNR value when comparing signals $\textbf{x}^m$ with $\textbf{r}_f^m$ according to~\eqref{eq:snr_val}. 

The values of $\delta^m$ are given by~\eqref{eq:delta_eq} and show that for instances where $\alpha^m = 0$, the value of $\delta_0$ is assigned to $\delta^m$ conversely, where $\alpha = 1$ the value of $\delta^m$ is given as $\delta_1$.

\begin{equation}
\label{eq:delta_eq}
    \delta^m = 
    \begin{cases}
         \delta_0   & \text{if} \; \alpha^m = 0\\
         \delta_1   & \text{if} \; \alpha^m = 1\\
    \end{cases}
\end{equation}

\begin{equation}
\label{eq:delta_0}
    \delta_0 = \frac{1}{M}\sum\limits_{m=1}^{M}(\delta_0^m)
\end{equation}

\begin{equation}
\label{eq:delta_1}
    \delta_1 = mode(\delta_1^m) \;\; \forall{m}~.
\end{equation}

Equation~\eqref{eq:delta_eq} and~\eqref{eq:delta_0} show that $\delta^m$ is equal to $\delta_0$, being the average value of $\delta^m_0$ for all values of $m$, and it can be seen from~\eqref{eq:delta_eq} and~\eqref{eq:delta_1} that $\delta^m$ is is equal to $\delta_1$, being the modified mode function $mode(\delta^m_1)$ that returns the most frequent element of $\delta^m_1$ for all values of $m$. Given the condition where multiple values appear equally as frequently in $\delta^m_1$, the $mode(\delta^m_1)$ function returns a randomly selected $\delta^m_1$ from the subset of most frequent appearing values. For the case where all values of $\delta^m_1$ appear equally as frequently, signifying that $M \leq \Theta$, the modified mode function returns an randomly chosen value of $\delta^m_1$.  

Indeed in subsection~\ref{subsec:model_training}, for convenience, the labelling algorithm details operations for one signal $z(k)$ of length $K$ reshaped into a matrix of $M$ windowed signals of length $N$, however, it should be understood that the complete model is trained on multiple noisy signals reshaped into windowed waveforms, further details are presented in section~\ref{subsec:experimenta_setup}.

\subsection{Classification Algorithms}\label{subsec:class_model}

The task of identifying the optimum filtering method between a low-pass elliptical filter and a wavelet filter can be formulated as a binary classification problem. As such, various machine learning models, including DNN models, can be applied to this problem. In this paper, we propose a convolution neural network (CNN) classification model for the given task. The CNN model is chosen due to its ability to learn unknown variations in the input distribution in the input data \cite{goodfellow2016deep}, such as noise. This model is compared against other machine learning models, such as support vector machines (SVM), logistic regression, K-nearest neighbours (KNN) and a DNN.

\begin{figure*}[ht]
    \centering
]    \includegraphics[width=0.91\textwidth]{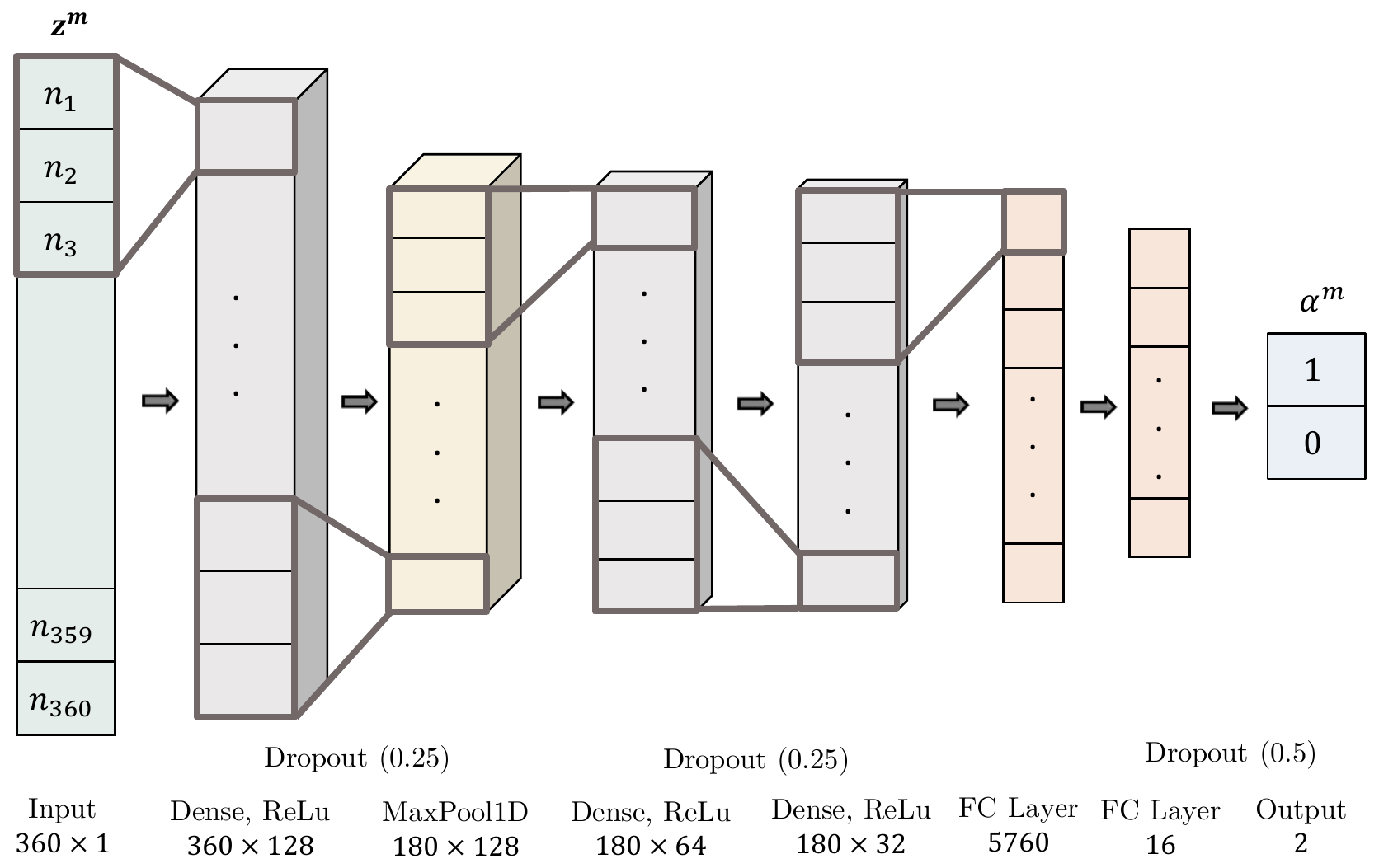}
    \caption{The DNN was configured with the following parameters in sequential order: a 128 dense layers, a maximum pooling layer with pooling size of 2, a 64 dense layers, a 32 dense layers and a 16 dense layers, all using ReLu activation functions. A unit dropout rate of $25\%$ was used after the 128 dense layer and 64 dense layer, followed by a $50\%$ rate after the flattened 16 dense layer, this was applied in order to avoid overfitting \cite{srivastava14a}. An stochastic gradient decent (SGD) optimizer using back propagation was used for the learning method and the model was trained for 20 epochs.}
    \label{fig:training_algo}
\end{figure*}

The hyperparameters for DNN is be shown in Fig.~\ref{fig:training_algo}, which details the number of layers, activation functions, pooling layers and flattening layers used. To summarise, a 128 dense layer, followed by a max-pooling layer of size 2 with an equal stride value with a stride value, and a 64 dense layer were added, the dense layer using rectified linear unit (ReLu) activation functions. This is followed by a 32 dense layer and a flattened 16 layer, all using the ReLu activation. The ReLu activation function was chosen based on fine tuning of the model on the training dataset. The final layer consisted of 2 dense layers using a \textit{SoftMax} activation function. This model was configured with a kernel size of 3. The optimisation function chosen for this model was Adam: a method for stochastic optimization~\cite{kingma2014adam} with parameters: learning rate = $0.001$, $\beta_1 = 0.9$, $\beta_2 = 0.999$, $\epsilon = 1e-07$. A categorical cross-entropy function was used to calculate loss and the model was trained for $20$ epochs of time and a batch size of 16 was used.

\subsection{Feature Reduction Techniques}
\label{subsec:feature_reduction}

Two different feature reduction techniques, principal components analysis (PCA) and independent analysis (ICA), were used for the models evaluated for this task. Reducing the dimensions of the data being considered, particularly using techniques that retain any significant information from the data such as PCA and ICA, have shown to be effective in improving classification accuracies in tasks involving ECG signal analysis \cite{elhaj2016arrhythmia}.

When applying PCA, the dimension of the data was reduced by $95\%$ such that only the top $5\%$ of the principal components were used. For ICA, 36 independent components were used, being $10\%$ of the initial signal length. These parameters for PCA and ICA were chosen to retain the most important features whilst reducing the length of the data, thus improving the performance of the classification models. 

\subsection{Experimental Set-up}
\label{subsec:experimenta_setup}

For the classification task at hand only the filter labels $\alpha^m$ from $y^m$ for a windowed noisy signal $\textbf{z}^m$ are required for the filter classification model. The parameter $\delta^m$ is used for selecting the data being applied to the models and $\beta^m$ is used to assign the optimum filter with an appropriate filter parameter. Particularly when calculating $\delta^m$, the complete dataset of windowed signals should be used after construction, such that the value of $M$ used for~\eqref{eq:delta_0} and~\eqref{eq:delta_1} is replaced by the total number of windowed signals for all full signals, shown in~\eqref{eq:multipl_sigs}.

Using the elliptical filter, given by~\eqref{eq:ellip_eq} and~\eqref{eq:ellip_psi}, requires the presetting of parameters such as the filter order, passband ripple and stopband ripple. For this study, a 7-th-order filter with a $3 \; \text{dB}$ passband and $4 \; \text{dB}$ stopband was chosen. These parameters control the elliptical filter response characteristics and were chosen based on prior testing of signals that returned the best average performance by the filter.

The value of $\beta^m$ for a windowed signal is used to remove any signal anomalies; there are cases where clean signals from the dataset are corrupted, such that the original waveform shows zero-amplitude and in some cases the clean signal itself exhibits noise. We require an ideal, clean signals in order to control the variables of the study, thus we choose remove signals where the underlying waveforms are distorted. It was observed, from prior analysis of the training data, that instances where signals are corrupted have a $\beta^m$ value less than $-3.00 \; \text{dB}$ approximately, thus signals where $\beta^m \leq -3.00 \; \text{dB}$ are removed from the dataset being used on the models.

Detailed in~\ref{subsec:model_training}, the signal reshaping and labeling method is specified for one noisy signal $z(k)$ reshaped into a matrix $Z$ of windowed signals. The proposed models are trained and tested on multiple noisy signals $z^l(k)$, generated from multiple different clean signals $x^l(k)$, where $l = \{1,2,\ldots,358\}$ and the resultant noisy windowed waveforms are given as per: 

\begin{equation}
\label{eq:multipl_sigs}
     \text{Z}_{total} = [\{\text{Z}_1,\textbf{y}^T_1\},\{\text{Z}_2,\textbf{y}^T_2\}, \ldots, \{\text{Z}_{358},\textbf{y}^T_{358}\}]~.
\end{equation}
A $\lambda$ value of 360 was chosen, thus $M$ and $N$ values  in~\eqref{eq:signal_transform} are 10 and 360 respectively. For all 358 signals in our dataset, the total training sets of signal matrix $\text{Z}$ is given in~\eqref{eq:multipl_sigs} which adds up to $3580 $ labelled waveforms (examples of noisy signals. For our classification models, we split the data into training and test sets respectively of sizes 67\% and 33\%, which gives us 2399 windowed waveforms in training set and 1181 windowed waveforms in the test set.

\section{Results and Discussions}
\label{sssec:results}

\begin{figure}[h]
    \centering
    \includegraphics[width=0.48\textwidth]{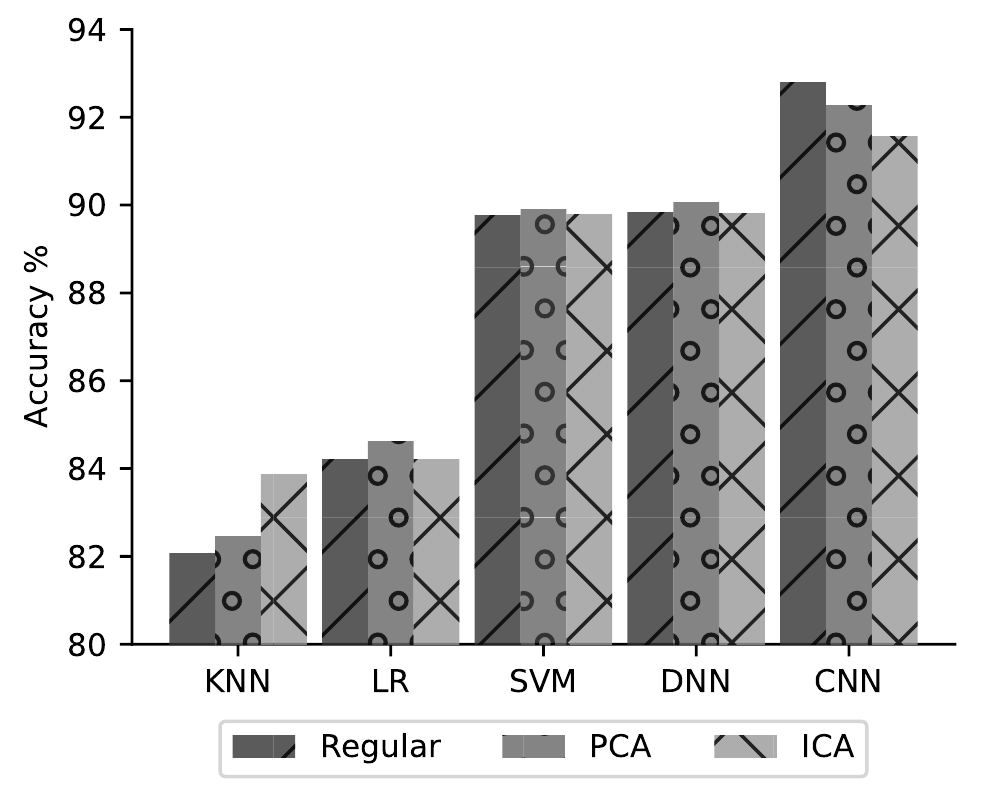}
    {\caption{Classification accuracies of SVM, KNN, logistic regression, DNN and CNN for the task of classifying the optimum filtering method for a given windowed signal.}
    \label{fig:accuracy_chart}}
\end{figure}

Table~\ref{tab:class_accuracies} lists the performance (classification accuracies) of classifiers. It can be seen from Fig.~\ref{fig:accuracy_chart} that SVM, logistic regression and DNN showed an improved model performance using only the principal components, compared to using the full data or independent components only. The CNN model achieved the highest classification accuracy for all three forms of input data; regular signal, principal components and independent components. %A drawback to consider for using the CNN network over other similarly performing methods is the computational complexity of the model itself, which due to the size of the proposed model, is not overly high with respect to other models. 

%\subsection{Model evaluation}

\begin{table}[ht]
    \centering
    {\caption{Classification accuracies of filter selection classifier. Models tested include SVM, logistic regression, KNN, DNN and CNN. Learning models were applied to the regular dataset and dataset after having applied feature reduction techniques such as PCA and ICA, see~\ref{subsec:feature_reduction}.}
        \label{tab:class_accuracies}}
    \begin{tabular}{lr}
    \toprule
    Classification model & Classification accuracy (\%)\\
    \midrule
        SVM &  89.77\\
        Logistic regression & 84.20\\
        KNN & 82.06\\
        DNN & 89.82\\
        CNN & \textbf{92.80}\\
        \\
        SVM (PCA) &  89.89\\
        Logistic regression (PCA) & 84.62\\
        KNN (PCA) & 82.23\\
        DNN (PCA) & 90.05\\
        CNN (PCA) & 92.27\\
        \\
        SVM (ICA) &  89.78\\
        Logistic regression (ICA) & 84.20\\
        KNN (ICA) & 84.46\\
        DNN (ICA) & 89.81\\
        CNN (ICA) & 91.56\\
    \bottomrule
    \end{tabular}
\end{table}

\subsection{Experimental Results}\label{subsec:experimental_res}

The results of a test signal $\textbf{z}^a$, representing a noisy signal with an arbitrary index $a$, such that $a \in m$ and is applied to both the wavelet filter and elliptical filters; this can be observed from Table~\ref{tab:ellip_signal_comp} along with their corresponding waveforms in Fig.~\ref{fig:ellip_sigs}. Comparatively, the filter responses for a signal $\textbf{z}^b$, where $b$ is an arbitrary signal index such that $b \in m$ and $a \neq b$, is shown in Fig.~\ref{fig:wavelet_sigs} with its corresponding data displayed in Table~\ref{tab:ellip_signal_comp}. It should be noted that the two signals considered, $\textbf{z}^a$ and $\textbf{z}^b$, both result in differing optimum filtering methods. 

\subsection{Discussion}
\label{subsec:discussion}

\begin{table}[h]
    \centering
    {\caption{Comparing the wavelet filter and elliptical filter on two signals: $\textbf{x}^a$ and $\textbf{x}^b$. Where the signals
    $\textbf{r}^a_0$ and $\textbf{r}^b_0$ represent the elliptical filtered responses for $\textbf{z}^a$ and $\textbf{z}^b$ respectively. The filtered signals $\textbf{r}^a_1$ and $\textbf{r}^b_1$ represent the wavelet filtered responses of $\textbf{z}^a$ and $\textbf{z}^b$ respectively. Showing the SNR and RMSE values of the noisy signal, signal filtered by wavelets and signal filtered through elliptical filtering.}
            \label{tab:ellip_signal_comp}}
    \begin{tabular}{l@{\hskip 0.5in}c@{\hskip 0.5in}r@{\hskip 0.5in}}
    \toprule
    Signal & RMSE & SNR (dB) \\
    \midrule
    \vspace{0.75ex}
    $\textbf{x}^a$                & -                 & - \\
    \vspace{0.75ex}
    $\textbf{z}^a$                & 0.154             & 6.846\\
    \vspace{0.75ex}
    % $R_0^a(\textbf{z}^a, \beta_0^a)$   & \textbf{0.122}    & \textbf{8.505} \\
    $\textbf{r}^a_0$   & \textbf{0.122}    & \textbf{8.505} \\
    \vspace{0.75ex}
    % $R_1^a(\textbf{z}^a, \beta_1^a)$    & 0.138             & 7.809 \\
    $\textbf{r}^a_1$    & 0.138             & 7.809 \\
    
    \\
    
    \vspace{0.75ex}
    $\textbf{x}^b$                & -                 & - \\
    \vspace{0.75ex}
    $\textbf{z}^b$                & 0.125             & 8.914 \\
    \vspace{0.75ex}
    % $R_0^b(\textbf{z}^b, \beta_0^b)$    & 0.122             & 9.591  \\
    % \vspace{0.5ex}
    $\textbf{r}^b_0$    & 0.122             & 9.591  \\
    \vspace{0.75ex}
    % $R_1^b(\textbf{z}^b, \beta_1^b)$    & \textbf{0.112}    & \textbf{10.024} \\
    $\textbf{r}^b_1$    & \textbf{0.112}    & \textbf{10.024} \\
    \bottomrule
    \end{tabular}
\end{table}

\begin{figure}[h]
    \centering
    \includegraphics[width=0.48\textwidth]{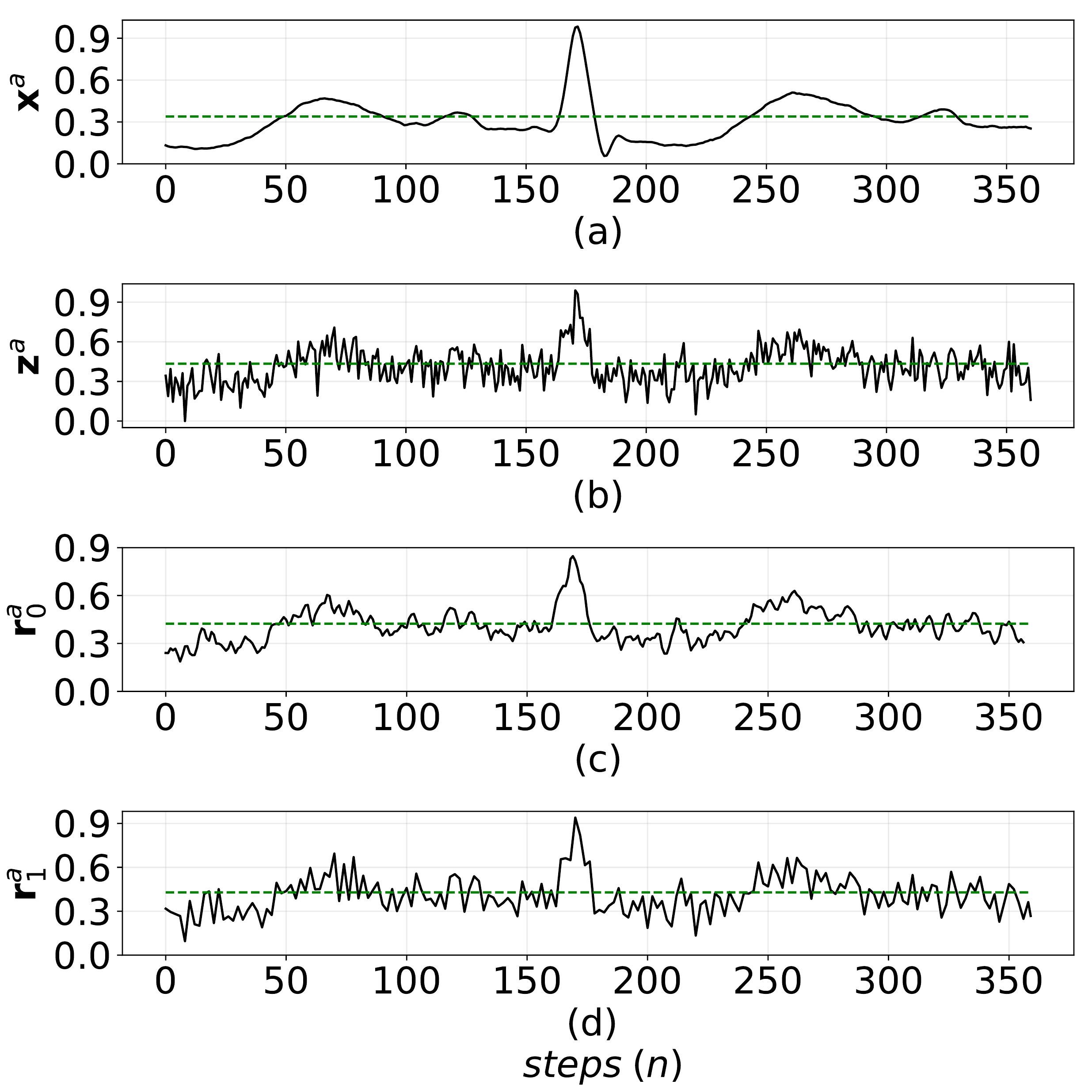}
    \caption{Signals shown where the optimum filter is determined to be the elliptical filter ($\alpha^a = 0$). Showing the clean signal $\textbf{x}^a$ (a), the signal with additive Gaussian noise applied to it $\textbf{z}^a$ (b), the noisy signal after having been processed through an optimum elliptical filter $\textbf{r}^a_0$ (c), and finally the noisy signal after having been filtered by an optimum wavelet filter $\textbf{r}^a_1$ (d). Outlined in black are the target waveforms and the green dotted lines represent the RMS values of each signal.}
    \label{fig:ellip_sigs}
\end{figure}

It can be observed from Table~\ref{tab:ellip_signal_comp} where the SNR and RMSE values, given by~\eqref{eq:SNR_eq} and~\eqref{eq:rmse_eq}, for noisy signals $\textbf{z}^a$ and $\textbf{z}^b$ are presented with their corresponding filtered signals. The signals $\textbf{r}^a_0 = R_0(\textbf{z}^a, \delta_0)$ and $\textbf{r}^a_1 = R_1(\textbf{z}^a, \delta_1)$ are the elliptical and wavelet filtered responses respectively for the noisy signal $\textbf{z}^a$. Whereas the signals $\textbf{r}^b_0 = R_0(\textbf{z}^b, \delta_0)$ and $\textbf{r}^b_1 = R_1(\textbf{z}^b, \delta_1)$ represent the elliptical and wavelet filtered signals respectively for noisy signal $\textbf{z}^b$.

\begin{figure}[h]
    \centering
    \includegraphics[width=0.48\textwidth]{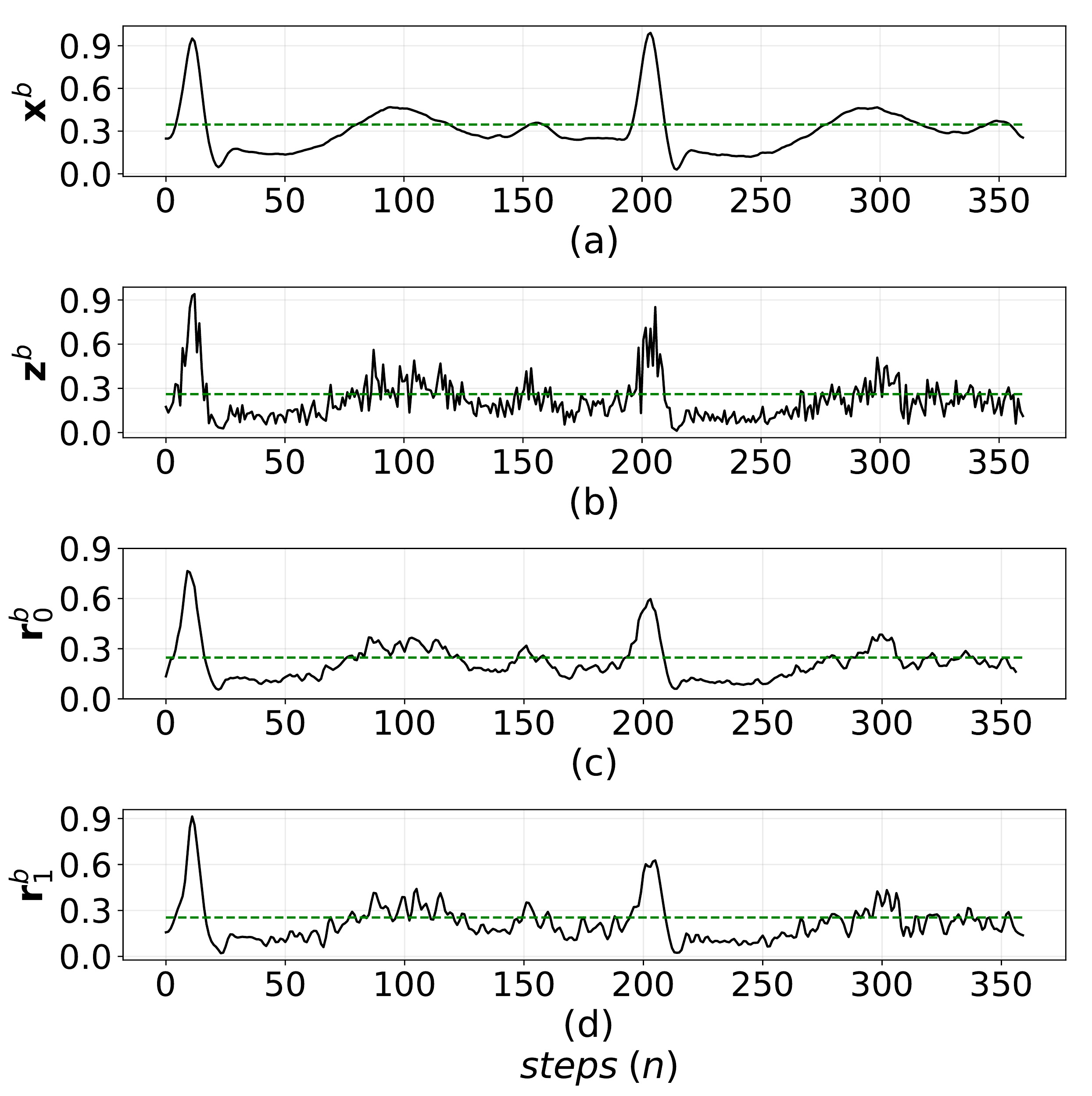}
    \caption{Signals shown where the optimum filter is determined to be the wavelet filter ($\alpha^b=1$). Showing the clean signal $\textbf{x}^b$ (a), the signal with additive Gaussian noise applied to it $\textbf{z}^b$ (b), the noisy signal after having been processed through an optimum elliptical filter $\textbf{r}^b_0$ (c), and finally the noisy signal after having been filtered by an optimum wavelet filter $\textbf{r}^b_1$ (d). Outlined in black are the target waveforms and the green dotted lines represent the RMS values of each signal.}
    \label{fig:wavelet_sigs}
\end{figure}

For noisy signal $\textbf{z}^a$, the lowest RMSE and highest SNR values were obtained using the elliptical filter response $\textbf{r}^a_0$. Contrastingly, given the noisy signal $\textbf{z}^b$, the lowest RMSE and highest SNR values were given by the filtered signal $\textbf{r}^b_1$, being the wavelet filter response. This proves that for a noisy signal, the successful classification of an optimum filter would indeed result in a response with a higher SNR value and lower RMSE value compared to the alternative filtering method. The results for which presented in Table~\ref{tab:ellip_signal_comp} and thus reaffirm the understanding that different windowed noisy signals ultimately result in differing optimum filtering methods, this can be further noticed from Figure.~\ref{fig:ellip_sigs} and Figure.~\ref{fig:wavelet_sigs}. Such a finding can be attributed to various signal characteristics; the two waveforms presented show differences in noise levels, baseline characteristics and average signal power, which all show to affect the optimum filtering method used for the signal.

From Table~\ref{tab:class_accuracies} it can be found that feature reduction methods, such as PCA and ICA, do not necessarily result in the highest classification accuracy for all models. This is evident particularly for the proposed CNN model, where the best classification performance was obtained with the original dataset, without having applied any feature reduction methods. Reduction of the dataset dimension should be  dismissed entirely, as the DNN model performance in Fig.~\ref{fig:accuracy_chart} shows that PCA being used as the input data resulted in the highest classification accuracy for that model. As stated by~\cite{Pepe_2020}, tasks involving signals with high levels of noise, where filter parameters are to be determined by a learning model, require fine tuning and experimentation. The findings in this paper show that it is possible to develop a DNN model to recognise noisy signals based on their optimum predicted filtering method. 

This investigation is an exploratory venture into developing and applying deep learning techniques to tasks involving digital signal processing, specifically for the selection of digital filtering methods. For the task of removing noise from ECG signals we investigate the wavelet filter, as proposed by various studies carried out for similar applications~\cite{zhang2019ecg, Shikhsarmast_2018, kabir2012denoising, POUNGPONSRI2013206}, and an elliptical filter, due to its frequency response at the defined cut-off frequency~\cite{Haykin1996Adaptive, Joshi2013ASurvey}. Consequently, various avenues of research are yet to be explored, such as using novel complex networks capable of processing complex signals and developing graphical representations of signals to be used is conjunction with machine learning models.

\section{Conclusions}
\label{ssssec:conclusion}

We propose a deep convolutional neural network (CNN) architecture for classifying an optimum signal denoising filter for a given noisy ECG signal. Moreover, we introduce an algorithm for labelling of signal waveforms with the optimum denoising filters (elliptical filter and wavelet filter). Our three versions of labelled datasets (full features dataset and reduced features datasets based on principal component analysis and independent component analysis) was fed to various classifiers such as, support vector machine, K-nearest neighbour, logistic regression and deep neural network and CNN. Our CNN model was able to classify the optimum filter with an accuracy of 92.8\%  when using the full feature dataset. Such a high classification accuracy for determining the optimum denoising filter enables us to effectively remove noise from a signal without affecting the underlying signal characteristics. We show that when presented with different windowed signals, the optimum filter can be either the elliptical filter (shown in Figure.~\ref{fig:ellip_sigs}) or the wavelet filter (shown in Figure~\ref{fig:wavelet_sigs}).

\bibliographystyle{IEEEtran}
\bibliography{References}

\begin{thebibliography}{10}
\providecommand{\url}[1]{#1}
\csname url@samestyle\endcsname
\providecommand{\newblock}{\relax}
\providecommand{\bibinfo}[2]{#2}
\providecommand{\BIBentrySTDinterwordspacing}{\spaceskip=0pt\relax}
\providecommand{\BIBentryALTinterwordstretchfactor}{4}
\providecommand{\BIBentryALTinterwordspacing}{\spaceskip=\fontdimen2\font plus
\BIBentryALTinterwordstretchfactor\fontdimen3\font minus
  \fontdimen4\font\relax}
\providecommand{\BIBforeignlanguage}[2]{{%
\expandafter\ifx\csname l@#1\endcsname\relax
\typeout{** WARNING: IEEEtran.bst: No hyphenation pattern has been}%
\typeout{** loaded for the language `#1'. Using the pattern for}%
\typeout{** the default language instead.}%
\else
\language=\csname l@#1\endcsname
\fi
#2}}
\providecommand{\BIBdecl}{\relax}
\BIBdecl

\bibitem{KHOKHAR20151650}
S.~Khokhar, A.~A. B.~M. Zin], A.~S.~B. Mokhtar, and M.~Pesaran, ``A
  comprehensive overview on signal processing and artificial intelligence
  techniques applications in classification of power quality disturbances,''
  \emph{Renewable and Sustainable Energy Reviews}, vol.~51, pp. 1650 -- 1663,
  2015.

\bibitem{paulson2005ultra}
C.~N. Paulson, J.~T. Chang, C.~E. Romero, J.~W. M.D., F.~J. Pearce, and N.~L.
  M.D., ``{Ultra-wideband radar methods and techniques of medical sensing and
  imaging},'' in \emph{Smart Medical and Biomedical Sensor Technology III},
  B.~M. Cullum and J.~C. Carter, Eds., vol. 6007, International Society for
  Optics and Photonics.\hskip 1em plus 0.5em minus 0.4em\relax SPIE, 2005, pp.
  96 -- 107.

\bibitem{Nassif2019Speech}
A.~B. {Nassif}, I.~{Shahin}, I.~{Attili}, M.~{Azzeh}, and K.~{Shaalan},
  ``Speech recognition using deep neural networks: A systematic review,''
  \emph{IEEE Access}, vol.~7, pp. 19\,143--19\,165, 2019.

\bibitem{Foster2014MachineLM}
K.~R. Foster, R.~Koprowski, and J.~D. Skufca, ``Machine learning, medical
  diagnosis, and biomedical engineering research - commentary,''
  \emph{BioMedical Engineering OnLine}, vol.~13, pp. 94 -- 94, 2014.

\bibitem{moody2001impact}
G.~B. Moody and R.~G. Mark, ``The impact of the {MIT-BIH} arrhythmia
  database,'' \emph{IEEE Engineering in Medicine and Biology Magazine},
  vol.~20, no.~3, pp. 45--50, 2001.

\bibitem{proakis2007digital}
J.~Proakis and D.~Manolakis, \emph{Digital Signal Processing}.\hskip 1em plus
  0.5em minus 0.4em\relax Pearson Prentice Hall, 2007.

\bibitem{Haykin1996Adaptive}
S.~Haykin, \emph{Adaptive Filter Theory (3rd Ed.)}.\hskip 1em plus 0.5em minus
  0.4em\relax USA: Prentice-Hall, Inc., 1996.

\bibitem{kovac2013adaptive}
B.~Kova{\v{c}}evi{\'c}, Z.~Banjac, and M.~Milosavljevi{\'c}, \emph{Adaptive
  Digital Filters}.\hskip 1em plus 0.5em minus 0.4em\relax Springer Berlin
  Heidelberg, 2013.

\bibitem{Joshi2013ASurvey}
S.~L. {Joshi}, R.~A. {Vatti}, and R.~V. {Tornekar}, ``A survey on {ECG} signal
  denoising techniques,'' in \emph{2013 International Conference on
  Communication Systems and Network Technologies}, 2013, pp. 60--64.

\bibitem{huikuri2001sudden}
H.~V. Huikuri, A.~Castellanos, and R.~J. Myerburg, ``Sudden death due to
  cardiac arrhythmia's,'' \emph{New England Journal of Medicine}, vol. 345,
  no.~20, pp. 1473--1482, 2001.

\bibitem{MARTIS201211792}
R.~J. Martis, U.~R. Acharya, K.~Mandana, A.~Ray, and C.~Chakraborty,
  ``Application of principal component analysis to {ECG} signals for automated
  diagnosis of cardiac health,'' \emph{Expert Systems with Applications},
  vol.~39, no.~14, pp. 11\,792--11\,800, 2012.

\bibitem{antczak2018deep}
K.~Antczak, ``Deep recurrent neural networks for {ECG} signal denoising,''
  \emph{arXiv preprint arXiv:1807.11551}, 2018.

\bibitem{acharya2017application}
U.~R. Acharya, H.~Fujita, S.~L. Oh, Y.~Hagiwara, J.~H. Tan, and M.~Adam,
  ``Application of deep convolutional neural network for automated detection of
  myocardial infarction using {ECG} signals,'' \emph{Information Sciences},
  vol. 415, pp. 190--198, 2017.

\bibitem{li2013advanced}
J.~Li, L.~Liu, Z.~Zeng, and F.~Liu, ``Advanced signal processing for vital sign
  extraction with applications in {UWB} radar detection of trapped victims in
  complex environments,'' \emph{IEEE Journal of Selected Topics in Applied
  Earth Observations and Remote Sensing}, vol.~7, no.~3, pp. 783--791, 2013.

\bibitem{lazaro2010analysis}
A.~Lazaro, D.~Girbau, and R.~Villarino, ``Analysis of vital signs monitoring
  using an {IR-UWB} radar,'' \emph{Progress In Electromagnetics Research}, vol.
  100, pp. 265--284, 2010.

\bibitem{liang2018ultra}
X.~Liang, J.~Deng, H.~Zhang, and T.~A. Gulliver, ``Ultra-wideband impulse radar
  through-wall detection of vital signs,'' \emph{Scientific Reports}, vol.~8,
  no.~1, pp. 1--21, 2018.

\bibitem{Shikhsarmast_2018}
F.~Shikhsarmast, T.~Lyu, X.~Liang, H.~Zhang, and T.~Gulliver, ``Random-noise
  denoising and clutter elimination of human respiration movements based on an
  improved time window selection algorithm using wavelet transform,''
  \emph{Sensors}, vol.~19, no.~1, p.~95, Dec 2018.

\bibitem{kabir2012denoising}
M.~A. Kabir and C.~Shahnaz, ``Denoising of {ECG} signals based on noise
  reduction algorithms in {EMD} and wavelet domains,'' \emph{Biomedical Signal
  Processing and Control}, vol.~7, no.~5, pp. 481--489, 2012.

\bibitem{POUNGPONSRI2013206}
S.~Poungponsri and X.-H. Yu, ``An adaptive filtering approach for
  electrocardiogram {(ECG)} signal noise reduction using neural networks,''
  \emph{Neurocomputing}, vol. 117, pp. 206 -- 213, 2013.

\bibitem{wang2020ecg}
G.~Wang, L.~Yang, M.~Liu, X.~Yuan, P.~Xiong, F.~Lin, and X.~Liu, ``{ECG} signal
  denoising based on deep factor analysis,'' \emph{Biomedical Signal Processing
  and Control}, vol.~57, p. 101824, 2020.

\bibitem{Pepe_2020}
G.~Pepe, L.~Gabrielli, S.~Squartini, and L.~Cattani, ``Designing audio
  equalization filters by deep neural networks,'' \emph{Applied Sciences},
  vol.~10, no.~7, p. 2483, Apr 2020.

\bibitem{Jenkal2016499}
W.~Jenkal, R.~Latif, A.~Toumanari, A.~Dliou, O.~El~B'Charri, and
  F.~Maoulainine, ``An efficient algorithm of {ECG} signal denoising using the
  adaptive dual threshold filter and the discrete wavelet transform,''
  \emph{Biocybernetics and Biomedical Engineering}, vol.~36, no.~3, pp.
  499--508, 2016.

\bibitem{ramirez2001performance}
F.~Ramirez-Mireles, ``On the performance of ultra-wideband signals in gaussian
  noise and dense multipath,'' \emph{IEEE Transactions on Vehicular
  Technology}, vol.~50, no.~1, pp. 244--249, 2001.

\bibitem{zhang2019ecg}
D.~Zhang, S.~Wang, F.~Li, J.~Wang, A.~K. Sangaiah, V.~S. Sheng, and X.~Ding,
  ``An {ECG} signal de-noising approach based on wavelet energy and sub-band
  smoothing filter,'' \emph{Applied Sciences}, vol.~9, no.~22, p. 4968, 2019.

\bibitem{goodfellow2016deep}
I.~Goodfellow, Y.~Bengio, and A.~Courville, \emph{Deep Learning}.\hskip 1em
  plus 0.5em minus 0.4em\relax MIT Press, 2016.

\bibitem{kingma2014adam}
D.~P. Kingma and J.~Ba, ``Adam: A method for stochastic optimization,''
  \emph{arXiv preprint arXiv:1412.6980}, 2014.

\bibitem{srivastava14a}
N.~Srivastava, G.~Hinton, A.~Krizhevsky, I.~Sutskever, and R.~Salakhutdinov,
  ``Dropout: A simple way to prevent neural networks from overfitting,''
  \emph{Journal of Machine Learning Research}, vol.~15, no.~56, pp. 1929--1958,
  2014.

\bibitem{elhaj2016arrhythmia}
F.~A. Elhaj, N.~Salim, A.~R. Harris, T.~T. Swee, and T.~Ahmed, ``Arrhythmia
  recognition and classification using combined linear and nonlinear features
  of {ECG} signals,'' \emph{Computer Methods and Programs in Biomedicine}, vol.
  127, pp. 52 -- 63, 2016.

\end{thebibliography}

\end{document}